\documentclass[twocolumn,twoside,slac_two]{revtex4}
\usepackage{graphicx}
\usepackage{epsfig}
\usepackage{fancyhdr}
\usepackage{amsmath,amssymb}
\newcommand{\ncred}{\mbox{$ \sigma_{r,{\rm NC}}^{\pm}$}}

\newcommand{\ncdd}{\mbox{$\frac{\textstyle {\rm d^2} \sigma^{\pm}_{{\rm NC}}}{\textstyle {\rm d}x{\rm d} Q^2}$}}

\newcommand{\ccdd}{\mbox{$\frac{\textstyle {\rm d^2} \sigma^{\pm}_{{\rm CC}}}{\textstyle {\rm d}x{\rm d} Q^2}$}}

\newcommand{\ccred}{\mbox{$ \sigma_{r,{\rm CC}}^{\pm}$}}
\newcommand{\ccredp}{\mbox{$ \sigma_{r,{\rm CC}}^{+}$}}
\newcommand{\ccredm}{\mbox{$ \sigma_{r,{\rm CC}}^{-}$}}

\newcommand{\Fig}{\mbox{figure}}

\newcommand{\Eq}{\mbox{equation}}

\newcommand{\FFig}{\mbox{Figure}}

\newcommand{\Figs}{\mbox{figures}}

\newcommand{\Eqs}{\mbox{equations}}

\newcommand{\FFigs}{\mbox{Figures}}

\newcommand{\bs}{\overline{s}}
\newcommand{\bc}{\overline{c}}
\newcommand{\bu}{\overline{u}}
\newcommand{\bd}{\overline{d}}

\newcommand{\bU}{\overline{U}}
\newcommand{\bD}{\overline{D}}

\def\dof{\mathop{n_{\rm dof}}\nolimits}

\begin{document}

\title{Recent Results from HERA}

%

\author{S. Glazov}
\affiliation{DESY 
  Notkestrasse 85, 
        22607 Hamburg, Germany}

\begin{abstract}
HERA $ep$ collider provides unique information on the proton structure. 
High center of mass energy $s=320$~GeV gives access to  both the low Bjorken-$x$
domain and regime of high momentum transfers $Q$. Recently the H1 collaboration
reported a high precision measurement of the 
structure function $F_2$  at low $x$
leading to tight constraints on the sea quark and gluon densities.
Both the H1 and ZEUS collaborations measure the structure function $F_L$ which 
provides an important cross check of the conventional QCD picture. This
measurement is recently extended by H1 to low $Q^2$ where small $x$  
corrections may play important role. An ultimate precision
of the deep inelastic scattering cross section measurement is achieved
by combining the measurements of the H1 and ZEUS collaborations.
The combined data are used as a sole input to a QCD fit to obtained
HERA PDF set. 
New measurements of inclusive $e^-p$ neutral and charged current scattering cross sections by the ZEUS collaboration at high $Q^2$ improve precision in this kinematic domain. 
H1 analysis of the DIS high $P_t$ jet production cross section is used for a determination
of the strong coupling constant $\alpha_S$. Separation of the strange quark density from the total sea
is obtained by the HERMES collaboration using tagged $K^{\pm}$ production. 
\end{abstract}

\maketitle

\thispagestyle{fancy}


\section{Introduction}
Deep inelastic lepton-hadron scattering (DIS) is important for the understanding
of the structure of the proton and of the dynamics of parton interactions.
The discovery of Bjorken scaling~\cite{Bloom:1969kc}
and its violation~\cite{Fox:1974ry} at fixed target experiments 
triggered the development
of the theory of strong interactions, Quantum Chromodynamics (QCD).
Significant progress in the exploration of strong interactions
has been achieved at the electron-proton collider HERA.  

The high center-of-mass energy of the $ep$ scattering 
at HERA leads to  a wide kinematic range
extending to large values of the absolute of the four-momentum transfer squared,
 $Q^2$, and to very small values of the Bjorken $x$ variable.
At the HERA beam energies of $E_e=27.6$\,GeV for the electron 
and $E_p=920$\,GeV for the proton, Bjorken $x$ values as small
as $10^{-4}$ ($10^{-6}$) are accessible for $Q^2$ of $10$\,GeV$^2$
($0.1$\,GeV$^2$) and $Q^2$ values as high as $30000$\,GeV$^2$ for high $x$.

HERA operation spanned over $15$ years, from $1992$ until $2007$, with a shutdown in $2000-2002$ to upgrade luminosity and instal spin rotators, to enable longitudinal beam polarization for the colliding experiments. The two colliding experiments, H1 and ZEUS, collected the DIS data for the whole HERA running period.
The data before the luminosity upgrade is used for high precision measurements at low $Q^2$ and low $x$, where the scattering cross section is high. The data after the luminosity upgrade is focused more on high $Q^2$ analyzes and polarization studies. For the last three months of its operation HERA run at reduced proton beam energy to measure the  proton structure function $F_L$.  

This paper is organized as follows. The DIS cross-section formulae and kinematic reconstruction at HERA are discussed in Sec~\ref{sec:kin}. The resent measurements of the DIS cross section at low $Q^2$ and QCD analysis of these data performed by the H1 collaboration are presented in Sec~\ref{sec:h1low}. An ultimate experimental precision is achieved by combination of the H1 and ZEUS data which is presented in Sec~\ref{sec:comb}. These combined data are used as input for a QCD fit, termed HERAPDF0.2. The measurement of the structure function $F_L$ performed by both H1 and ZEUS collaboration is presented in Sec~\ref{sec:fl}. The new measurement of the neutral and charge 
current scattering cross-sections are given in Sec~\ref{sec:nch2} and Sec~\ref{sec:cch2}. H1 measurement of the 
DIS jet-production cross-section and determination of the strong coupling constant $\alpha_S$ is described in Sec~\ref{sec:alphas}. The flavor decomposition of the parton densities using data collected by H1 and ZEUS is given in Sec~\ref{sec:flavour}.

\section{DIS Cross Section  \label{sec:kin}}
\begin{figure*}[htb]
\vspace*{1cm}
\centerline{
\epsfig{file=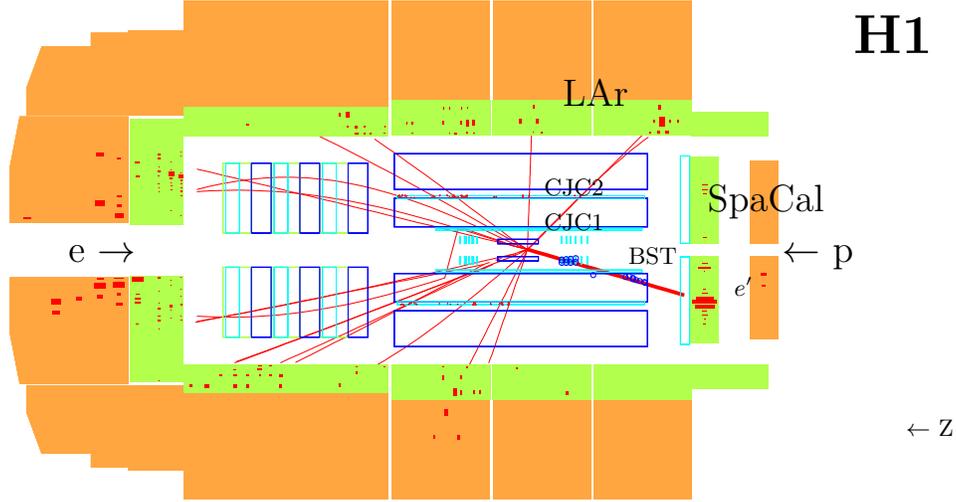,angle=90,width=0.5\linewidth}
}
\vspace*{-0.25cm}\hspace*{-12cm}\begin{picture}(10,5)
\put(30,56){ \Large  e $ \rightarrow $}
\put(300,56){ \Large  $ \leftarrow $ p}
\put(210,68){ CJC1}
\put(210,81){ CJC2}
\put(350,-10){$ \leftarrow$  Z}
\put(220,115){\Large LAr}
\put(275,75){\Large SpaCal}
\put(285,43){$e'$}
\put(245,55){BST}
\put(330,135){\huge \bf H1}
\end{picture}
\vspace*{1cm}
\caption{A view of a high $y$ event reconstructed in the H1 detector.
The positron and proton beam directions are indicated by the arrows.
For the coordinate system used at HERA the $z$ axis points in the direction
of the proton beam.
The interaction vertex is reconstructed using the hadronic final state
(thin lines) and the scattered positron (thick line) tracks in the central
tracker. The central tracker consists of (from the beam line outwards)
the silicon tracker, the drift chambers CJC1 and CJC2, it is surrounded
by the liquid argon (LAr) calorimeter. The detector operates in a solenoidal
magnetic field of $1.16$~T.
The scattered 
positron trajectory is reconstructed
in the backward silicon tracker BST and the CJC1. 
The charge of the particle is determined using the track curvature.
The positron energy
is measured in the electromagnetic part of the SpaCal calorimeter. 
\label{fig:event}}
\end{figure*}
The neutral  current deep inelastic $e^{\pm}p$ scattering cross section, at tree level,
is given by a linear combination of generalized structure functions. For unpolarized beams
it can be expressed as
\begin{eqnarray} \label{ncsi}     
 \ncred =\ncdd \cdot \frac{Q^4 x}{2\pi \alpha^2 Y_+}                                                     
  =            \tilde{F}_2 \mp \frac{Y_-}{Y_+} \tilde{xF}_3 -\frac{y^2}{Y_+} \tilde{F}_L,
\end{eqnarray}                                                                  
where the electromagnetic coupling, the photon              
propagator and a helicity factor are absorbed 
in the definition of a reduced cross section \ncred, and $Y_{\pm}=1 \pm (1-y)^2$.                                                                              
The functions $\tilde{F_2}$, $\tilde{F_L}$  and $\tilde{xF_3}$
depend on  the electroweak parameters as:
\begin{eqnarray} \label{strf}                                                   
 \tilde{F_2} &=& F_2 - \kappa_Z v_e  \cdot F_2^{\gamma Z} +                      
  \kappa_Z^2 (v_e^2 + a_e^2 ) \cdot F_2^Z, \nonumber \\   
 \tilde{F_L} &=& F_L - \kappa_Z v_e  \cdot F_L^{\gamma Z} +                      
  \kappa_Z^2 (v_e^2 + a_e^2 ) \cdot F_L^Z, \nonumber \\                     
 \tilde{xF_3} &=&  \kappa_Z a_e  \cdot xF_3^{\gamma Z} -                     
  \kappa_Z^2 \cdot 2 v_e a_e  \cdot xF_3^Z.                                   
\end{eqnarray} 
Here $v_e$ and $a_e$ are the vector and axial-vector weak couplings of 
the electron to the $Z$ boson 
and $\kappa_Z(Q^2) =   Q^2 /[(Q^2+M_Z^2)(4\sin^2 \Theta \cos^2 \Theta)]$. 
At low $Q^2$, the contribution of $Z$
 exchange is negligible and $\sigma_{r,{\rm NC}} = F_2  - y^2 F_L/Y_+$.
The contribution of the term containing the structure function $F_L$ is only significant for large values of $y$.

In the Quark Parton Model (QPM)  
the  structure function $F_L$ is zero\,\cite{PhysRevLett.22.156} and the 
other functions in \Eq\,\ref{strf} are given as
\begin{eqnarray} \label{ncfu}                                                   
  (F_2, F_2^{\gamma Z}, F_2^Z) &=&  [(e_u^2, 2e_uv_u, v_u^2+a_u^2)(xU+ x\bar{U}) \nonumber \\
  & + &   (e_d^2, 2e_dv_d, v_d^2+a_d^2)(xD+ x\bar{D})],            
                                 \nonumber \\                                   
  (xF_3^{\gamma Z}, xF_3^Z) &=& 2x  [(e_ua_u, v_ua_u) (xU-x\bar{U}) \nonumber \\
  & + &   (e_da_d, v_da_d) (xD-x\bar{D})] ,                        
\end{eqnarray} 
where  $e_u,e_d$ denote the electric charge of up- or
down-type quarks while $v_{u,d}$ and $a_{u,d}$ are 
the vector and axial-vector weak couplings of the up- or 
down-type quarks to the $Z$ boson.
Here   $xU$, $xD$, $x\bU$ and $x\bD$ denote
the sums of up-type, of down-type and of their 
anti-quark distributions, respectively. 
Below the $b$ quark mass threshold,
these sums are related to the quark distributions as follows
\begin{equation}  \label{ud}
\begin{array}{l}
  xU  = xu + xc,    ~~~~~
 x\bU = x\bu + x\bc, \\
  xD  = xd + xs,    ~~~~~
 x\bD = x\bd + x\bs\,, 
\end{array}
\end{equation}
where $xs$ and $xc$ are the strange and charm quark distributions.
Assuming symmetry between sea quarks and anti-quarks, 
the valence quark distributions result from 
\begin{equation} \label{valq}
xu_v = xU -x\bU, ~~~~~~~~~~ xd_v = xD -x\bD.
\end{equation}

A reduced cross section for 
the inclusive unpolarized charged current $e^{\pm} p$ 
scattering  is
\begin{equation}
 \label{Rnc}
 \ccred =  
  \frac{2 \pi  x}{G_F^2}
 \left[ \frac {M_W^2+Q^2} {M_W^2} \right]^2
          \ccdd.
\end{equation}
At leading order, the $e^+p$ and $e^-$ charged current scattering cross-sections are
\begin{eqnarray}
\label{ccupdo}
 \ccredp &=& x\bU+ (1-y)^2xD, \nonumber \\
 \ccredm &=& xU +(1-y)^2 x\bD . 
\end{eqnarray}
Therefore the NC and CC measurements may be used
to determine the combined sea quark distribution functions, $\bU$ and $\bD$,
and the valence quark distributions, $u_v$ and $d_v$. 
A QCD analysis in the DGLAP 
formalism~\cite{dglap}
also allows the 
gluon momentum distribution $xg$ in the proton 
to be determined from the scaling violations of the data.

For  NC scattering, event kinematics can be reconstructed using the scattered electron as well as 
the hadronic final state particles. A typical low $Q^2$, high $y$ event measured by the H1 detector is 
shown in \Fig~\ref{fig:event}. For  CC scattering, kinematics is reconstructed using the hadronic final state.
\section{Measurements of DIS Cross Section at low $\boldsymbol{Q^2}$ by H1 \label{sec:h1low}}
\begin{figure}
\epsfig{file=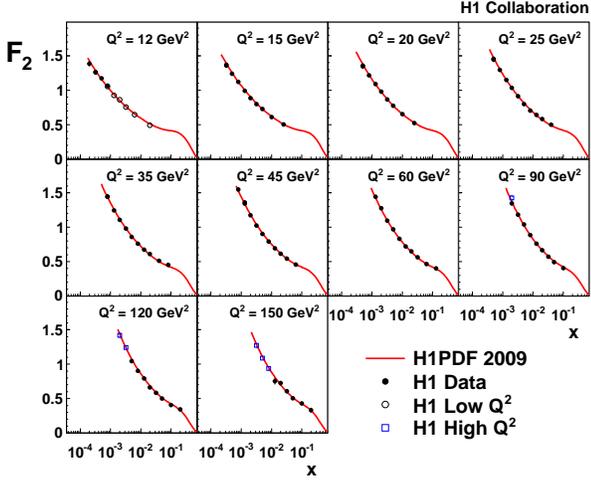,width=\linewidth}
\caption{
Measurement of the structure function
  $F_2$ at fixed $Q^2$ as a function of $x$ by the H1 collaboration.. 
The error bars represent
  the total measurement uncertainties. The curve represents the H1 PDF2009 QCD
  fit.
\label{fig:f2}
}
\end{figure}
\begin{figure}
\epsfig{file=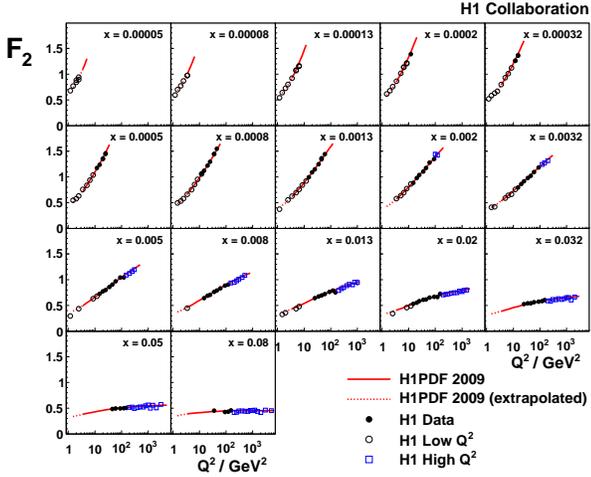,width=\linewidth}
\caption{
Measurement of the structure function
  $F_2$ as a function of $Q^2$ at various values of $x$ by the H1 collaboration.
The error bars represent
  the total measurement uncertainties. The solid curve represents the
  H1PDF 2009 QCD fit for $Q^2 \geq 3.5$~GeV$^2$, which is
  also shown extrapolated down to $Q^2 = 1.5$~GeV$^2$ (dashed).
\label{fig:f2q}}
\end{figure}
Recently the H1 collaboration reported new measurements of the NC cross section at low $0.2 \le Q^2\le 150$~GeV$^2$~\cite{h1a,h1b} based on data collected in 1999-2000.
This measurement comprises several analysis. The lowest $Q^2\le 1.5$~GeV$^2$ is reached using a dedicated HERA run in which the interaction vertex was shifted in the forward direction to allow detection of the electrons scattered at large angles.
For $1.5 \le Q^2\le 12$~GeV$^2$ the measurement is performed using a dedicated H1 run with modified trigger conditions.
Both measurements use the BST (HERA-I configuration) to reconstruct the scattered electron and measure the event vertex, and
employ events with initial state QED radiation from the electron to reach  lowest $Q^2$ values. The higher range in $12\le Q^2<150$~GeV$^2$ is measured under nominal beam and detector conditions. The event vertex is reconstructed by the CJC and the electron angle is measured using the vertex and the track element reconstructed in the backward drift chamber, which was mounted in front of the SpaCal for the HERA-I run period.

The measured structure function $F_2$ is shown in \Fig~\ref{fig:f2} as a function of $x$ for $12\le Q^2\le 150$~GeV$^2$. For each $Q^2$ bin, there is a strong rise of the structure functions for $x\to 0$, which indicates a large sea quark density.
The rate of the rise increases with increasing $Q^2$, which is a signature of a large gluon density. \FFig~\ref{fig:f2q} shows
the structure function $F_2$ for fixed values of $x$ as a function of $Q^2$. For low $x$, there is a strong $Q^2$ dependence of the structure function. For highest $x$, the data show approximate scaling behavior. In both  \Figs~\ref{fig:f2} and \ref{fig:f2q} the data is compared to an next-to-leading order (NLO) QCD fit, which describes data very well. This fit is discussed in the next section.
\subsection{H1PDF 2009 QCD Fit}
\begin{figure}
\begin{tabular}{lr}
\begin{minipage}{0.5\linewidth}
\epsfig{file=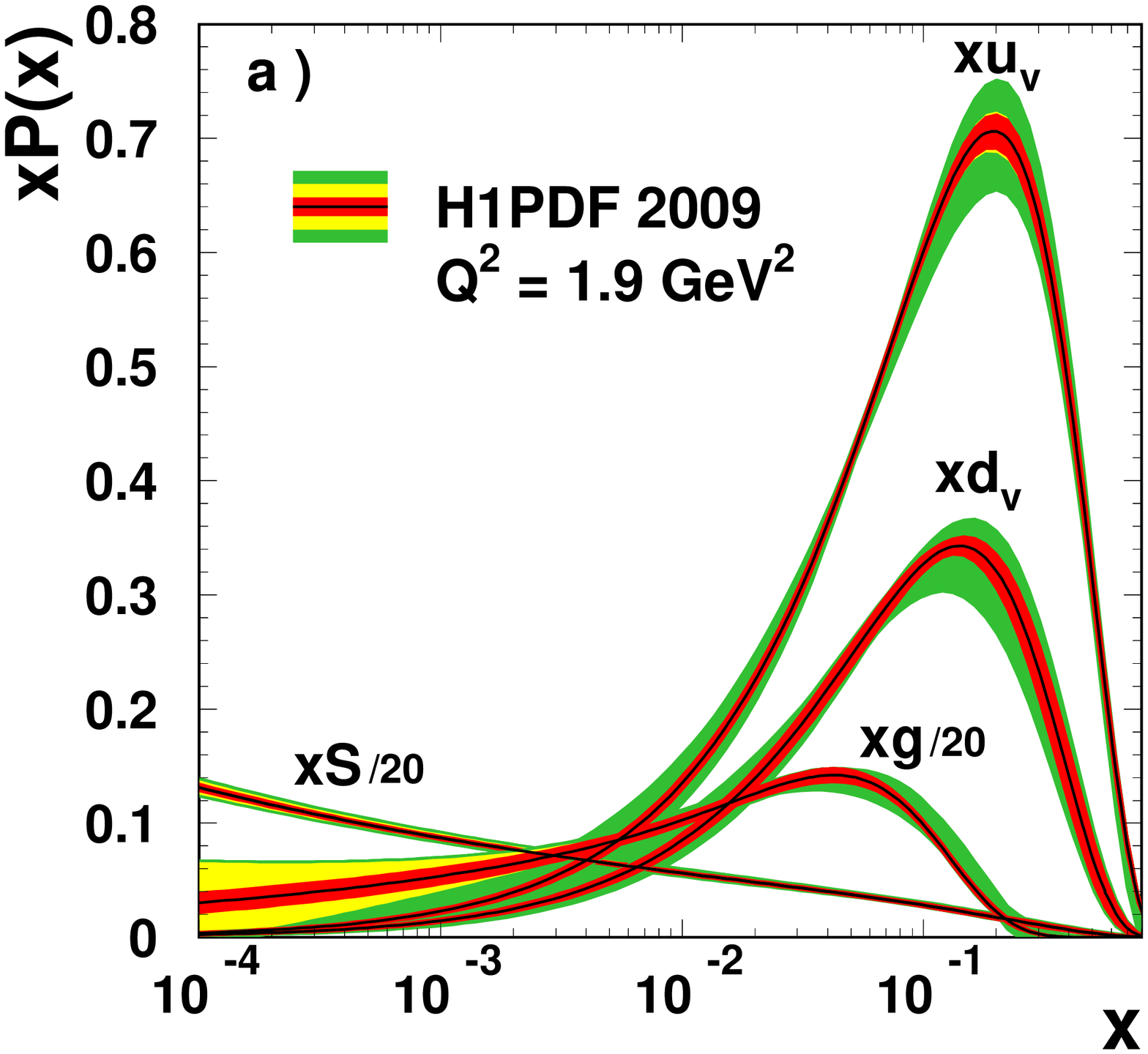,width=\linewidth}
\end{minipage} &
\begin{minipage}{0.5\linewidth}
\epsfig{file=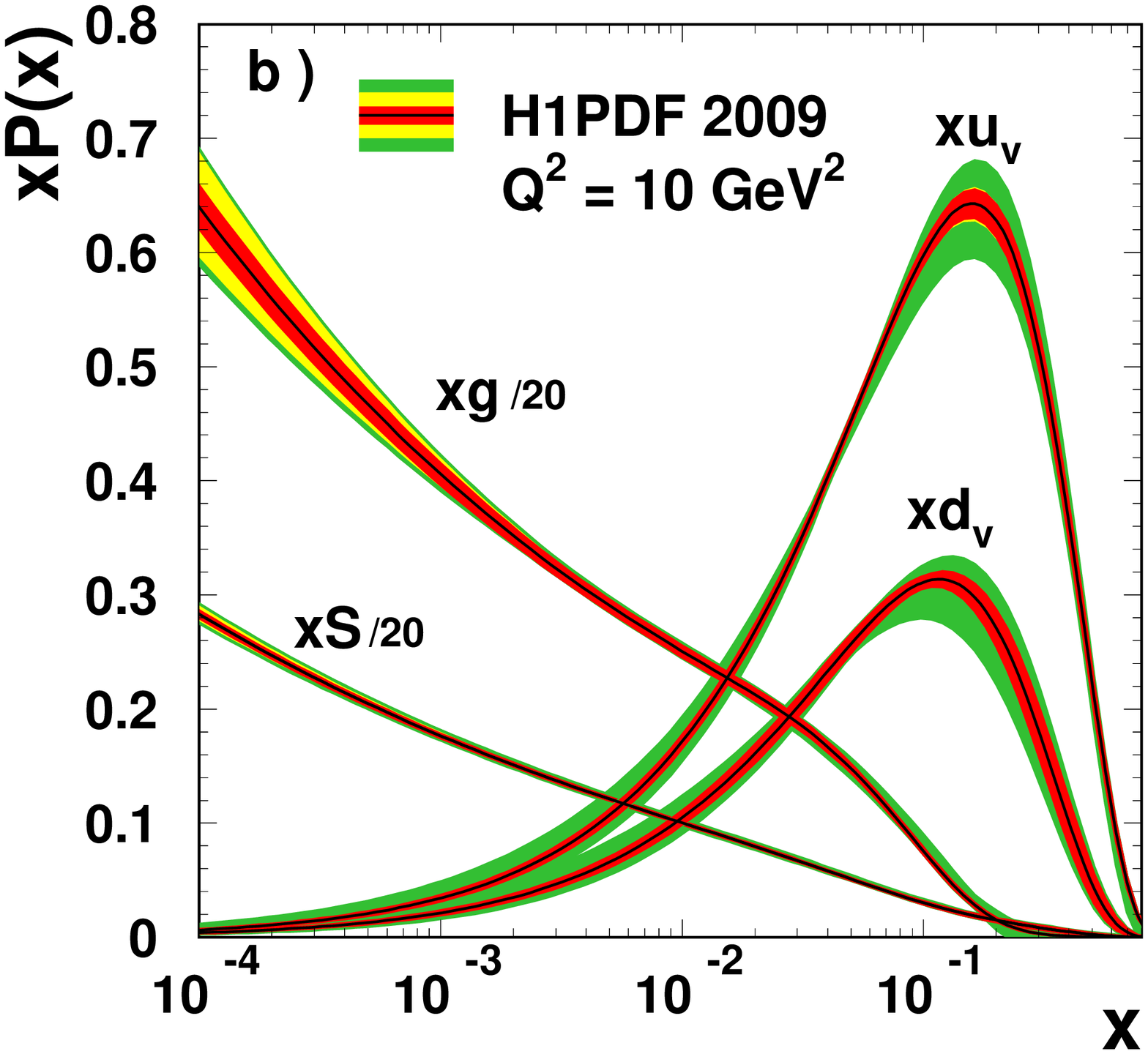,width=\linewidth}
\end{minipage}
\end{tabular}
\caption{Parton distributions as determined by
 the H1PDF 2009 QCD fit at $Q^2 = 1.9$~GeV$^2$ a) and at $Q^2 = 10$~GeV$^2$ b). 
In a) and c)  the gluon and sea-quark
 densities are divided by a factor $20$.
 The inner error bands show the experimental uncertainty, the middle
 error bands include the theoretical model uncertainties of the fit
 assumptions, and the outer error band represents the total
 uncertainty including the parameterization uncertainty.
\label{fig:h1pdf09}.}
\end{figure}
The new low $Q^2$ H1 data together with published previously H1 measurements of NC and CC $e^{\pm}p$ scattering cross sections~\cite{h1c} are used as a sole input in the NLO QCD analysis, termed H1PDF2009. The QCD evolution is performed the QCDNUM program~\cite{qcdnum}. The fit uses heavy-quark variable-flavor-number scheme of~\cite{rt1} with recent modifications~\cite{rt2}.

The parton densities are parameterized at the starting scale $Q^2_0=1.9$~GeV$^2$ using $xf(x) = Ax^B(1+x)^C(1+P_1 x+ P_2 x^2 + ...)$ form. Five PDFs are used to describe the H1 data: $xg$, $x\bar{U}$, $x\bar{D}$, $xu_v$ and $xd_v$. The QCD analysis imposes momentum and quark-number
sum rules, which determine $A_g$, $A_{u_v}$ and $A_{d_v}$. Starting from all polynomial coefficients $P_i=0$, the central
fit is chosen by adding one $P_i$ for one PDF at a time. The modified PDF is selected if (i) $\chi^2$ is significantly improved,
(ii) structure functions $F_2$ and $F_L$ are positive, (iii) all PDFs are positive, (iv) PDFs satisfy valence-dominance condition: $xu_v>x\bar{u}$ and $xd_v> x\bar{d}$ at high $x$. This analysis is repeated for all $i\le 3$.

The QCD analysis distinguishes three sources of PDF uncertainties: experimental, model and parameterization uncertainty.
The experimental uncertainty is evaluated using $\Delta \chi^2=1$ criterion. The model uncertainty is measured by modifying
the input parameters such as masses of the heavy quarks and the starting scale. The parameterization uncertainty is estimated
by adding more $P_i$ terms ($i\le 3$) and lifting the conditions (iii) and (iv).  The total uncertainty is given by a sum
in quadrature of the experimental, model and parameterization uncertainties.

The results of the fit are shown in \Fig.~\ref{fig:h1pdf09} for the starting scale $Q^2=1.9$~GeV$^2$ and for $Q^2=10$~GeV$^2$.
The experimental uncertainty is small for all PDFs at all values of $x$ compared to the total uncertainty. 
At large $x$, the valence quark densities have significant parameterization uncertainty. At low $x$, model uncertainty
plays bigger role.
\section{Combination of the H1 and ZEUS data \label{sec:comb}}
\begin{figure}
\epsfig{file=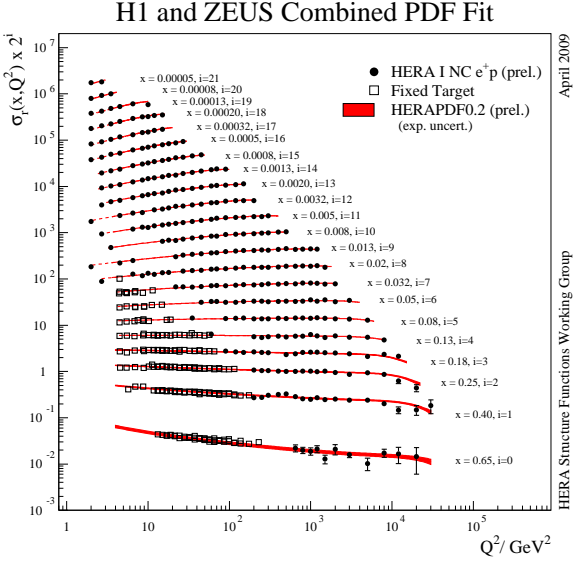,width=\linewidth}
\caption{\label{fig:scal}Measurement of the $e^+p$ NC scattering cross section based on the combination
of the H1 and ZEUS data compared to the fit to these data and measurements from fixed-target experiments.}
\end{figure}
The highest precision of the cross-sections measurements at HERA is obtained by combining the results from the H1 and ZEUS experiments. The combination of the results is performed taking into account the correlated systematic uncertainties~\cite{h1a,glaz}. The starting point is the $\chi^2$ function for individual measurement which is  defined as
\begin{equation}
\begin{array}{l}
 \chi^2_{\rm exp}\left(\boldsymbol{m},\boldsymbol{b}\right) =  \\
= \sum_i
 \frac{\left[m^i
- \sum_j \gamma^i_j m^i b_j  - {\mu^i} \right]^2}
{ \textstyle \delta^2_{i,{\rm stat}}\,{\mu^i}  \left(m^i -  \sum_j \gamma^i_j m^i b_j\right)+
\left(\delta_{i,{\rm uncor}}\,  m^i\right)^2} + \\
 + \sum_j b^2_j\,.
\label{eq:ave}
\end{array}
\end{equation} 
Here  ${\mu^i}$ is the  measured  value  at a point $i$ and
 $\gamma^i_j $, 
$\delta_{i,{\rm stat}} $ and 
$\delta_{i,{\rm uncor}}$ are relative
correlated systematic, relative statistical and relative uncorrelated systematic uncertainties,
respectively.
The function $\chi^2_{\rm exp}$ depends on the predictions $m^i$ 
for the measurements
(denoted as the vector $\boldsymbol{m}$) and 
 the shifts of correlated systematic error sources $b_j$ (denoted as $\boldsymbol{b}$).
For the reduced cross-section  measurements  ${\mu^i} = \sigma_r^i$,
$i$ denotes a $(x,Q^2)$ point, and the summation over
$j$ extends over all correlated systematic sources. 
The predictions $m^i$ are 
given by the assumption that there is a single true value of the cross section 
corresponding to each data point $i$ and each process, neutral or 
charged current $e^+p$ or $e^-p$ scattering.
Under the assumption that the statistical uncertainties are proportional
to the square root of the number of events and that the systematic
uncertainties are proportional to $\boldsymbol{m}$, the minimum of
 \Eq~\ref{eq:ave} provides an unbiased estimator of $\boldsymbol{m}$.

Several data sets providing a number of measurements are represented
by a total $\chi^2$ function,
which is built from the sum of the $\chi^2_{\rm exp}$ functions for each data set $e$
The data averaging procedure allows the rearrangement of the total $\chi^2$ such
that it takes a form similar to \Eq~\ref{eq:ave}.

The averaging procedure is applied to the H1 and ZEUS inclusive data from the HERA-I running period. 
All the NC and CC cross-section data from H1 and ZEUS are combined in one simultaneous minimization. 
Therefore resulting shifts of 
the correlated systematic uncertainties propagate coherently to both CC and NC data.
In total $1402$ data points are combined to $741$ cross-section measurements. 
The data show good consistency, with $\chi^2/\dof = 637/656$. 
In \Fig~\ref{fig:scal} the NC reduced cross section, for $Q^2 > 1$\,GeV$^2$, is shown
as a function of $Q^2$ for the 
HERA combined $e^+p$ data and for fixed-target data~\cite{bcdms,nmc} across 
the whole of the measured kinematic plane. The data precision reaches $1.1\%$ for $10\le Q^2\le 100$~GeV$^2$.

The combined data uses as a sole input for the NLO QCD analysis. The fit uses similar approach as the H1PDF2009 fit.
The result of the fit is compared to the data in \Fig~\ref{fig:scal}. There is a good 
agreement between the extrapolation of the fit to lower $Q^2$ and fixed-target data in this kinematic domain.
  
\section{Measurements of the Structure Function $\boldsymbol{F_L}$\label{sec:fl}}
\begin{figure}
\epsfig{file=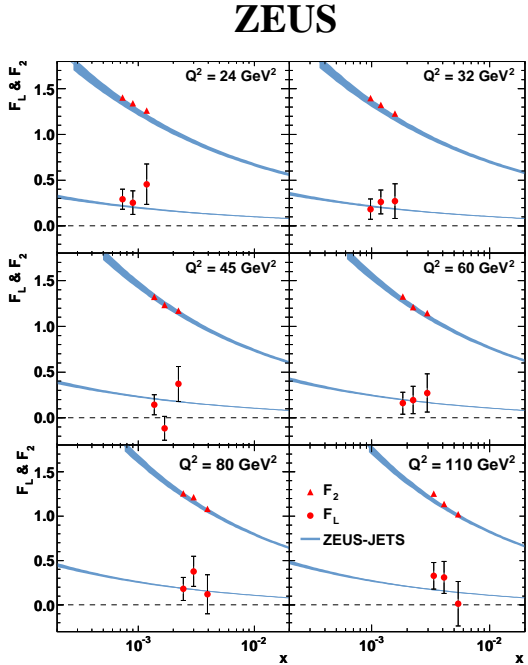,width=\linewidth}
\caption{Measurement of the structure functions $F_L$ and $F_2$ by the ZEUS collaboration compared to the prediction of ZEUS-JETS fit. \label{fig:flzeus}}.
\end{figure}
\begin{figure}
\epsfig{file=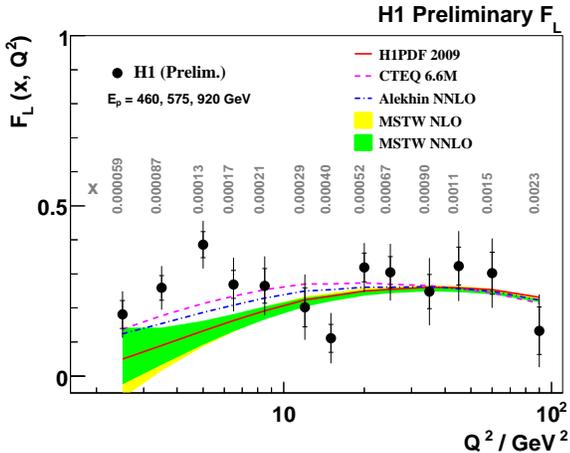,width=\linewidth}
\caption{Preliminary measurement of the structure functions $F_L$ and $F_2$ by the H1 collaboration. 
The data are quoted at fixed values of $Q^2$ and  $x$ indicated in grey\label{fig:flh1} and compared to predictions based on various models.}.
\end{figure}
The large gluon density, determined from scaling violation of the $F_2$ data using QCD fits (see \Fig~\ref{fig:h1pdf09}), implies
that the structure function $F_L$ must be significant at low $x$. Direct measurements of $F_L$ allow to test this prediction 
providing a check of the QCD  and adding an extra constraint for the gluon density.   

To determine the two structure functions $F_2(x,Q^2)$ and 
$F_L(x,Q^2)$ from the reduced
cross section
 it is necessary to perform measurements at the same
values of $x$ and $Q^2$ but different $y$. This is achieved at HERA by
reducing the proton beam energy. Two  $e^+p$ runs at
reduced proton beam energy   $E_p=460$~GeV and $E_p=575$~GeV were
performed with an integrated luminosity of about $13$~pb$^{-1}$ and 
$6$~pb$^{-1}$,
respectively. The run at $E_p=460$~GeV gives highest sensitivity to $F_L$ 
while the run at $E_p=575$~GeV extends the kinematic range of the measurement
and provides an important cross check.

The measurement must extend to as high $y$ as possible to increase sensitivity
to $F_L$. A high $y$ kinematic domain at low $Q^2$ corresponds to low energies
of the scattered positron $E'_e$. Measurement at low $E'_e$ is challenging
primarily because of high hadronic background. 

The ZEUS collaboration uses Monte Carlo (MC) simulation to estimate the background.
The MC prediction is normalized to the data using a sub-sample of tagged background events.
The results of the ZEUS analysis~\cite{flzeus} are shown in  \Fig.~\ref{fig:flzeus}. ZEUS report
measurement of both structure functions $F_2$ and $F_L$ obtained from a linear fit of the reduced
cross section as a function of $y^2/Y_+$. The structure function $F_L$ is found to be in a good agreement 
with the prediction of the ZEUS-JETS PDF set~\cite{zeusjet}.

To reduce and estimate
the hadronic background, H1 demand the scattered positron candidate to have a reconstructed
track with well measured  curvature. The curvature is used to determine the candidate charge.
For the signal, positive charge is expected. The background is approximately charge symmetric,
i.e. the number of background events with different charges is about equal. Using this, the  
background is estimated from the negative charge sample, corrected for a small charge asymmetry
and then subtracted from the positive charge sample. The charge asymmetry of the background is determined
directly from the data by comparing negatively and positively charged candidates from $e^+p$ and $e^-p$ 
data taking periods, respectively. Therefore, the background determination is purely data driven for the H1
analysis.

H1 published the first measurement of $F_L$ at HERA using drift chambers
CJC1 and CJC2 together with the SpaCal calorimeter
for $12\le Q^2\le 90$~GeV$^2$\cite{h1fl} and also reported a preliminary 
result at higher $Q^2$ using the LAr calorimeter~\cite{flh1comb}. 
These measurements were recently extended to lower $2.5<Q^2<12$~GeV$^2$ by using the backward silicon tracker BST~\cite{flbst}.

The H1 measurement of the structure function $F_L$ is shown in \Fig~\ref{fig:flh1}.
The data are compared to the predictions from various models. The predictions agree among each other
and with the data for $Q^2>10$~GeV$^2$. For lower $Q^2$, there is a notable difference between 
the NLO predictions of MSTW and CTEQ. This difference is traced down to the difference in accounting for
$\alpha^2_S$ corrections. The data are somewhat higher than both predictions and agree better with the  
CTEQ calculations.
\subsection{Neutral Current $\boldsymbol{e^-p}$ DIS Cross Section \label{sec:nch2}}
\begin{figure}
\epsfig{file=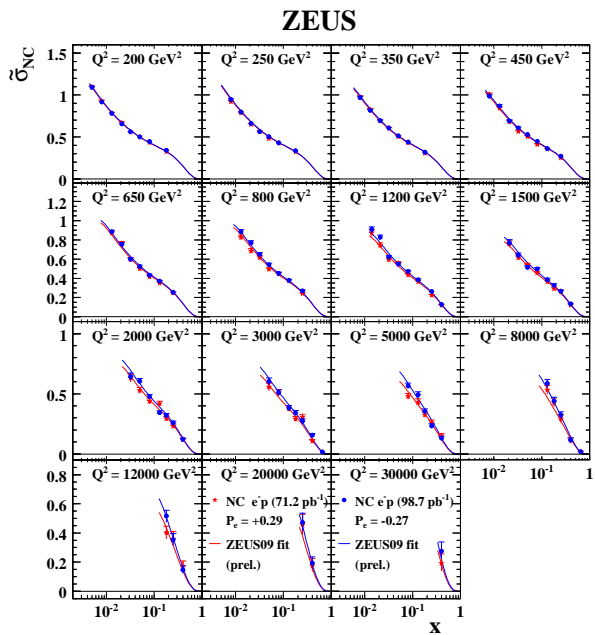,width=\linewidth}
\caption{NC $e^-p$ scattering cross section measured by the ZEUS collaboration. 
Data for positive (red stars) and negative (blue dots) polarization are compared
to the fit. 
\label{fig:zeusnc}}
\end{figure}
Measurements at high $Q^2$ using  data collected in years $2003$-$2007$ are published by 
the ZEUS collaboration. For the NC scattering, taking into account 
the longitudinal beam polarization and ignoring small contribution from pure $Z$ exchange,
the structure function $F_2$ is 
\begin{equation}
\tilde{F}_2 = F_2 - (v_e - P_e a_e) \kappa_Z F_2^{\gamma Z}\,,
\end{equation} 
where $P_e$ is the polarization value.
Therefore, there is a weak polarization dependence of the cross section which increases
at high $Q^2$. 

The ZEUS measurement of the NC $e^-p$ cross section for $P_e = +29\%$ and $P_e=-27\%$~\cite{zeusnc} is shown
in \Fig~\ref{fig:zeusnc}. The result is based on integrated luminosity 
of $169.9$~pb$^{-1}$ collected in 2005 and 2006.
The data are compared to a  preliminary QCD fit. At high $Q^2$, there
is an increasing difference between cross sections for the two polarizations.
\subsection{Charged Current $\boldsymbol{e^{\pm}p}$ DIS Cross Section \label{sec:cch2}}
\begin{figure}
\epsfig{file=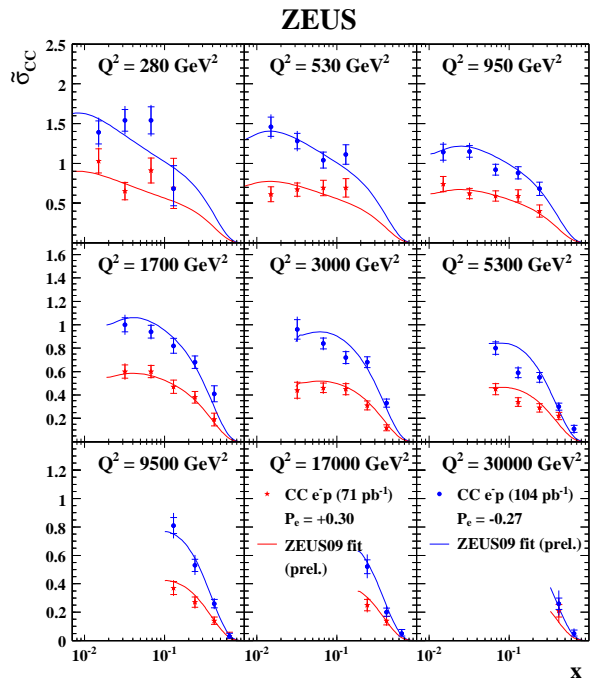,width=\linewidth}
\caption{CC $e^-p$ scattering cross section measured by the ZEUS collaboration. \label{fig:zeusccm}
Data for positive (red stars) and negative (blue dots) polarization are compared
to the fit. 
}
\end{figure}
\begin{figure}
\epsfig{file=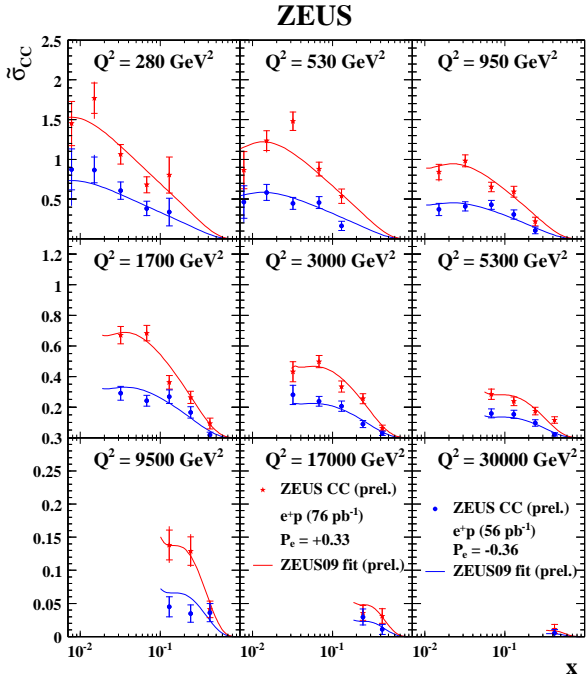,width=\linewidth}
\caption{Preliminary measurement of CC $e^+p$ scattering cross  by the ZEUS collaboration. \label{fig:zeusccp}
Data for positive (red stars) and negative (blue dots) polarization are compared
to the fit. 
}
\end{figure}
In the standard model, the CC cross section depends linearly on the longitudinal electron beam polarization. \FFigs~\ref{fig:zeusccm} and~\ref{fig:zeusccp} show published~\cite{zeuscc} and preliminary analyses of $e^-p$ and $e^+p$ 
CC cross sections performed by the ZEUS collaboration. 
For the published $e^-p$ sample, the same 2005-2006 data are used as for the
published NC sample, the preliminary $e^+p$ analysis is based on data collected
in 2003-2004 and 2006-2007. 
The data agree well with the expectations of the QCD fit. 
The $e^+p$ CC data are of additional value for the QCD analyses because they
are sensitive to the $d$ and $s$ quark densities which are less constraint
by the NC data, see \Eqs~\ref{ncfu} and ~\ref{ccupdo}.
\section{Jet Cross Section and Determination of $\boldsymbol{\alpha_S}$ \label{sec:alphas}}
\begin{figure}
\epsfig{file=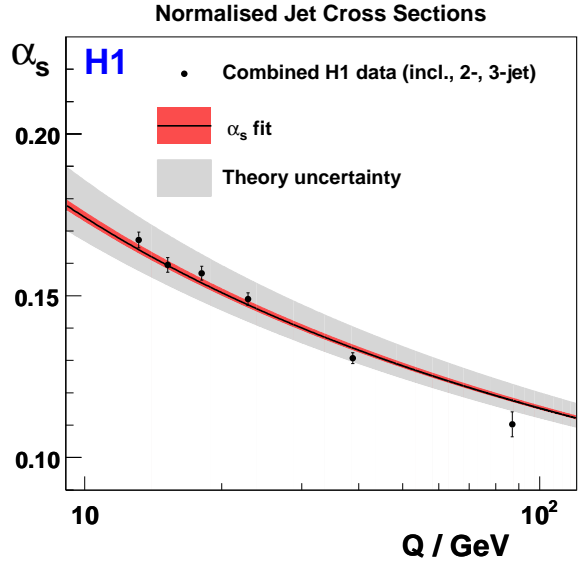,width=\linewidth}
\caption{The value of $\alpha_S(Q)$ obtained by H1 by a simultaneous fit
of all normalized jet cross sections.\label{alphafit}  } 
\end{figure}
The production of jets in DIS can be used to determine the gluon density and 
to measure the strong coupling constant $\alpha_S$. Recently H1 have perfomed
a measurement of inclusive, 2-jet and 3-jet cross sections~\cite{h1jets}
as a function of $Q^2$
using HERA data collected in 1999-2007 with an integrated luminosity of
395~$pb^{-1}$. The measurements are well described by NLO QCD calculations,
corrected for hadronization effects.

Using these data, H1 determine the strong coupling constant $\alpha_S$.
The measurement is performed separately for different $Q^2$ bins
and different processes. These measurements are combined
together and the evolution of $\alpha_S$ as a function of $Q$ is compared
to the theory prediction in \Fig~\ref{alphafit}. A good agreement is observed
between the data and theory. The theoretical uncertainty is dominated by the higher order corrections. The value of $\alpha_S$ at $M_Z$ 
\begin{equation}
\begin{array}{lcl}
\alpha_S(M_Z) &=& 0.1176 \pm 0.0020 (\mbox{exp.}) ^{+0.0046}_{-0.0030}(\mbox{th.}) \\
&\pm& 0.0016 (\mbox{PDF})
\end{array}
\end{equation}
agrees well with the world average. The experimental error on $\alpha_S$
is about $0.6\%$. The total uncertainty is dominated by the theory, it may be improved with calculation of NNLO corrections.
\section{Measurements of Strange, Charm and Bottom Quark Densities \label{sec:flavour}}
Inclusive NC cross section at low $Q^2$ and low $x$
 are dominated by the structure function $F_2$ and do not allow to separate contributions of individual quark flavors. The flavor separation can be achieved
in semi-inclusive scattering by tagging the struck quark flavor. For HERA
kinematics, tagging production of a charmed meson, e.g. $D^*$, almost certainly
corresponds to a scattering off a $c$-quark. Samples with 
secondary vertices are enriched with $c$ and $b$-quark scatterings.

\begin{figure}
\epsfig{file=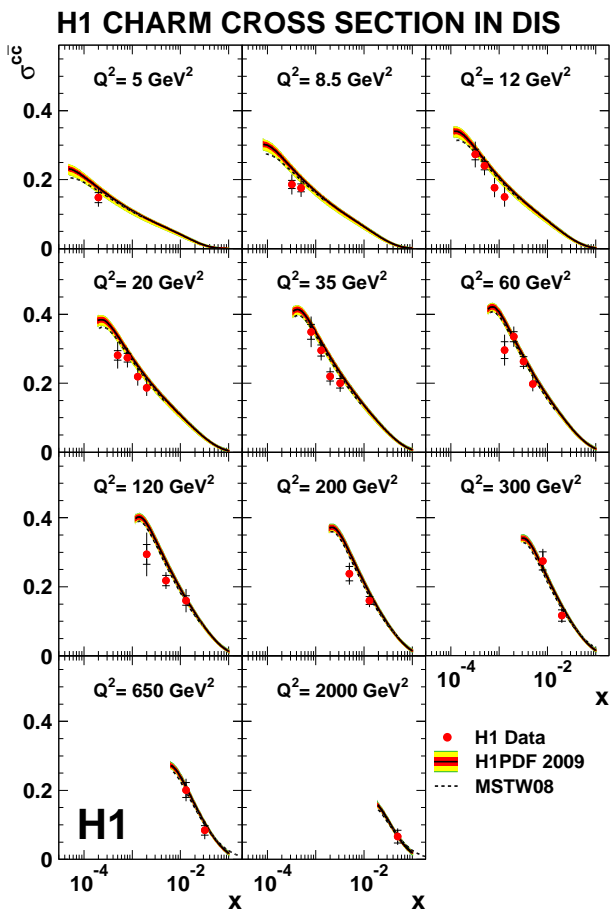,width=\linewidth}
\caption{\label{h1xc}The reduced cross section $\sigma^{c\bar{c}}$ measured
by H1 and shown as a function of $x$ for different $Q^2$ bins. The predictions 
of H1PDF 2009 and MSTW08 NLO fits to inclusive data are also shown.}
\end{figure}

\begin{figure}
\epsfig{file=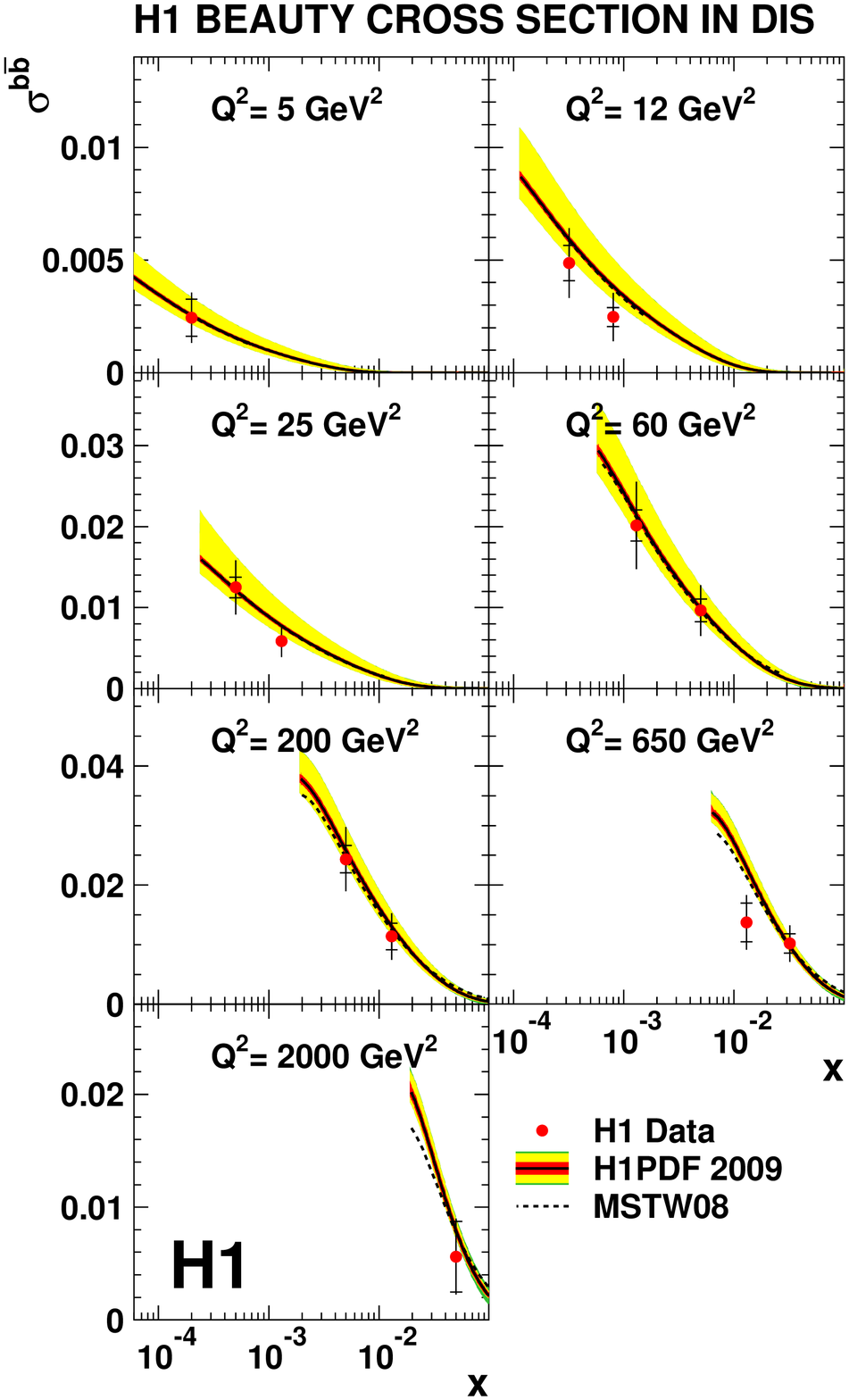,width=\linewidth}
\caption{\label{h1xb}The reduced cross section $\sigma^{b\bar{b}}$ measured
by H1 and shown as a function of $x$ for different $Q^2$ bins. The predictions 
of H1PDF 2009 and MSTW08 NLO fits to inclusive data are also shown.}
\end{figure}

To measure $b$ and $c$ structure functions using secondary vertices, 
it is essential to have high precision tracking detector installed close to the interaction point. Recently the H1 collaboration reported the measurement 
of $c$ and $b$ reduced cross sections~\cite{h1cb}, using secondary vertex 
technique,  based on complete sample for which  the central silicon tracker 
was installed. These data are shown in Figs~\ref{h1xc} and~\ref{h1xb} and compared to the H1PDF2009 fit. For all $Q^2$ bins there is a strong rise of the reduced cross section to low $x$ values which increases with increasing $Q^2$. This is a direct indication of the large gluon density. The data are in a good agreement with the H1PDF2009 fit. The uncertainty of the fit is dominated by the model
uncertainty due to the variation of the heavy quark masses. Therefore, these data allow to check the model and determine these parameters.  
\begin{figure}
\epsfig{file=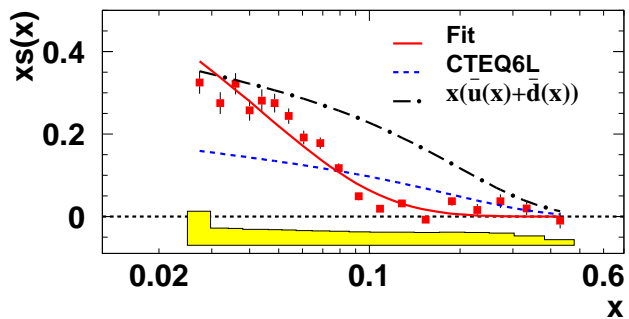,width=\linewidth}
\caption{\label{hermes}The strange parton distribution $xs(x)$
from the measured HERMES multiplicity for charged kaons evolved
to $Q^2_0=2.5$~GeV$^2$ . The solid curve is the HERMES fit to the
data, the dashed curve gives $xs(x)$ from the CTEQ6L set, and dot-dashed
curve us the sum of light antiquarks from the CTEQ6L set.}
\end{figure}

The fixed-target experiment HERMES, which operated using HERA $e^{\pm}$ beams, recently reported a measurement
the momentum distribution of the strange quark sea, $xs(x)$, in scattering off the deuteron. The strange density
is extracted from $K^{\pm}$ multiplicities by correcting for the $u,d$ and $s$ fragmentation. The result is shown
in \Fig~\ref{hermes} and compared to the CTEQ6L expectation. The HERMES data shows softer behavior of $xs(x)$ 
compared to the CTEQ6L fit: at high $x$ the data is below the fit and at low $x$ the data are above the fit.
At low $x$ the strange sea is comparable to the light sea density.
\section{Summary \label{sec:summary}}
Recent results from HERA provide new precise data for determination of the proton structure at low $x$. 
The combination of the H1 and ZEUS results reaches $1.1\%$ for the $10\le Q^2\le 100$~GeV$^2$ kinematic domain.
The data are well described by the NLO QCD fit. This fit has impressively small experimental uncertainties, the total uncertainties are dominated by the model and parameterization errors.

The conventional QCD picture is checked by the measurements of the structure function $F_L$ performed by H1 and ZEUS. Good agreement between the theory and the measurements is observed for $Q^2\ge 10$~GeV$^2$. For lower $2.5\le Q^2 < 10$~GeV$^2$, the preliminary data from H1 are somewhat above the expectations.

Measurements at high $Q^2$ using the polarized HERA-II data performed by ZEUS check the Standard Model and provide
constraints for the parton densities at high $x$. The flavor decomposition at high $x$ 
is achieved by using the charged current data.

The flavor decomposition at low $x$ and low $Q^2$ is performed by measuring semi-inclusive processes: tagged
heavy flavor production at H1 and $K^{\pm}$ production at HERMES. The H1 data for the charm and bottom quark
parton densities agree well with the predictions
from the QCD fit to the inclusive data. The HERMES data are softer than prediction of the CTEQ6L analysis.
  
Further results from HERA are expected as the data analysis are being finalized. In particular, combined H1 and ZEUS measurements of the inclusive HERA-II cross sections and combined measurement of the charm and bottom production  
will have significant impact on PDFs for the kinematic range important for the future measurements at the LHC.


\begin{thebibliography}{9}   
\bibitem{Bloom:1969kc} E.D. Bloom {\it et~al.}, Phys.\ Rev.\ Lett. 23 (1969) 930.
\bibitem{Fox:1974ry} D.J. Fox {\it et~al.}, Phys.\ Rev.\ Lett. 33 (1974) 1504.
\bibitem{PhysRevLett.22.156} C.G. Callan and D.J. Gross. Phys. Rev. Lett. {\bf 22}, 156 (1969).
\bibitem{dglap}  
V.N. Gribov and L.N. Lipatov, {Sov.\ J.\ Nucl.\ Phys.} 15 (1972) 438; \\ 
V.N. Gribov and L.N. Lipatov, {Sov.\ J.\ Nucl.\ Phys.} 15 (1972) 675; \\ 
L.N. Lipatov,  {Sov.\ J.\ Nucl.\ Phys.} 20 (1975) 94; \\ 
Y.L. Dokshitzer,   {Sov.\ Phys.\ JETP} 46 (1977) 641; \\
G. Altarelli and G. Parisi,  {Nucl.\ Phys.\ B} 126 (1977) 298.    
\bibitem{h1a} F. Aaron {\it et al}. [H1 Collaboration] (2009), [arXiv:0904.0929].
\bibitem{h1b} F. Aaron {\it et al}. [H1 Collaboration] (2009), [arXiv:0904.3513].
\bibitem{h1c} C. Adloff {\it et al}. [H1 Collaboration], Eur. Phys. J. C {\bf 30}, 1 (2003).
\bibitem{qcdnum} M. Botje, QCDNUM version 17{$\beta$}.
\bibitem{rt1} R.S. Thorne and R.G. Roberts, Phys. Rev. {\bf D57}, 6871 (1998). 
\bibitem{rt2} R.S. Thorne, Phys. Rev. {\bf D73}, 054019 (2006).
\bibitem{glaz} A. Glazov, AIP Conf. Proc. {\bf 792}, 237 (2005) [doi:10.1063/1.2122026].
\bibitem{bcdms} A. Benvenuti {\it et al.} [BCDMS Collaboration], Phys. Lett. {\bf B223}, 485 (1989).
\bibitem{nmc} M. Arneodo {\it et al.} [NMC Collaboration], Nucl. Phys. {\bf B483}, 3 (1997).
\bibitem{flzeus} S. Chekanov {\it et al.} [ZEUS Collaboration], Phys. Lett. {\bf B26255}, (2009).
\bibitem{zeusjet} S. Chekanov {\it et al.} [ZEUS Collaboration], Eur. Phys. J. {\bf C42}, 1 (2005).
\bibitem{h1fl}F.~Aaron {\it et al.} [H1 Collaboration], Phys.\ Lett.\
B 665 (2008) 139.
\bibitem{flh1comb} V. Chekelian [for the H1 Collaboration], 
 Proc.~of XVI Int.~Workshop on Deep-Inelastic Scattering and Related Topics, London, England, April 2008
\verb$doi: 10.3360/dis.2008.39$.
\bibitem{flbst} S. Glazov [for the H1 Collaboration], Proc.~of XVII Int.~Workshop on Deep-Inelastic Scattering
and Related Topics, Madrid, Spain, April 2009.
\bibitem{zeusnc} S. Chekanov {\it et al.} [ZEUS Collaboration], EPJ {\bf C62} 625 (2009).  
\bibitem{zeuscc} S. Chekanov {\it et al.} [ZEUS Collaboration], EPJ {\bf C61} 223 (2009).  
\bibitem{h1jets} F.D. Aaron {\it et al.} [H1 Collaboration], Submitted to EPJC, arxiv:0904.3870.
\bibitem{h1cb}   F.D. Aaron {\it et al.} [H1 Collaboration]. Accepted by EPJC. arxiv:0907.2643.
\bibitem{hermes}A. Airapetian {\it et al.} [HERMES Collaboration], Phys. Lett. {\bf B666} 446 (2008).

\end{thebibliography}
\end{document}